\documentclass{PoS}

\usepackage{epsfig}
\newcommand{\beqa}{\begin{eqnarray}}
\newcommand{\eeqa}{\end{eqnarray}}
\newcommand{\beq}{\begin{equation}}
\newcommand{\eeq}{\end{equation}}

\PoS{PoS(LAT2005)330}

\title{{\scriptsize \tt IPPP/05/59~~~DCPT/05/118}\\
\vspace{5mm}
Infrared exponents of Yang-Mills theory}

\ShortTitle{Infrared exponents of Yang-Mills theory}

\author{\speaker{Christian S. Fischer}\\
        IPPP, Durham\\
        E-mail: \email{christian.fischer@durham.ac.uk}}

\abstract{In this talk I summarise recent results on the infrared 
exponents of SU($N_c$)-Yang-Mills theory. I discuss a self-consistent power 
law solution for the Dyson-Schwinger equations for general 1PI-Greens 
functions in the infrared. The corresponding running coupling has a 
fixed point at zero momentum, which turns out to be universal and 
gauge invariant within a class of transverse gauges. When calculated 
on a torus the infrared exponents of the ghost and gluon propagators
differ from the corresponding continuum solutions. They agree, however,
well with results from lattice calculations.
}

\FullConference{XXIIIrd International Symposium on Lattice Field Theory\\
		 25-30 July 2005\\
		 Trinity College, Dublin, Ireland}

\begin{document}

The infrared behaviour of the Green's functions of SU($N_c$)-Yang-Mills theory
is related to confinement in several interesting ways. One example is the
Kugo-Ojima confinement criterion, which is satisfied in Landau gauge if the 
dressing function of the ghost propagator is singular and the gluon propagator 
is finite or vanishes in the infrared  \cite{Nakanishi:qmas,Alkofer:2000wg}. 
A second example is provided by the colour Coulomb potential, which can be 
determined from the instantaneous part of the time-time component of the 
Coulomb gauge gluon propagator \cite{Zwanziger:2003de}. In this talk I 
summarise recent analytical results on the infrared exponents of the Green's functions 
of SU($N_c$)-Yang-Mills theory in Landau gauge \cite{Alkofer:2004it} and a class 
of gauges that interpolate between Landau and Coulomb gauge \cite{Fischer:2005qe}.
Furthermore, I discuss the relation of these solutions with corresponding 
results from Dyson-Schwinger equations (DSEs) and lattice simulations on a 
compact manifold \cite{Torus}.

\section{Infrared exponents in the continuum: Landau gauge}

The basic idea to determine the infrared behaviour of one-particle-irreducible
(1PI) Green's functions is to investigate their Dyson-Schwinger equations 
order by order in a skeleton expansion ({\it i.e.} a loop expansion using 
full propagators and vertices).
It turns out that these Green's functions can only be infrared singular, if
all external scales go to zero. Thus to determine the degree of possible
singularities it is sufficient to investigate the
DSEs in the presence of only one external scale $p^2 \ll \Lambda^2_{QCD}$. 
In collaboration with R.~Alkofer and F.~Llanes-Estrada we 
showed by induction that the skeleton expansion is 
stable in the sense that higher orders generate terms that are
equally or less singular than those of lower orders. Thus one can 
determine the infrared behaviour of these functions from their DSEs 
approximated to lowest order in the skeleton expansion. 
We found a self-consistent solution of the tower of DSEs for 1PI-Green's functions,
given by \cite{Alkofer:2004it}
\beq
\Gamma^{n,m}(p^2) \sim (p^2)^{(n-m)\kappa}. \label{IRsolution}
\eeq
Here $\Gamma^{n,m}(p^2)$ denotes the infrared leading dressing function of 
the 1PI-Green's function with $2n$ external ghost legs and $m$ external 
gluon legs. The exponent $\kappa$ is known to be positive 
\cite{Watson:2001yv}, independent of any truncation scheme 
of the DSEs.  

A special instance of the solution (\ref{IRsolution}) are the inverse 
ghost and gluon dressing functions $\Gamma^{1,0}(p^2) = G^{-1}(p^2)$ and 
$\Gamma^{0,2}(p^2) = Z^{-1}(p^2)$, which are related to the ghost and
gluon propagators via
\beq
D^G(p^2) = - \frac{G(p^2)}{p^2} \, , \qquad
D_{\mu \nu}(p^2)  = \left(\delta_{\mu \nu} -\frac{p_\mu 
p_\nu}{p^2}\right) \frac{Z(p^2)}{p^2} \, .
\eeq
The corresponding power laws in the infrared are
\beq
G(p^2) \sim (p^2)^{-\kappa}, \hspace*{1cm} Z(p^2) \sim (p^2)^{2\kappa}\,.
\label{kappa}
\eeq
In this notation the Kugo-Ojima criterion translates to the condition
$\kappa \ge 0.5$. For a bare ghost-gluon vertex in the infrared, 
justified by lattice calculations \cite{Cucchieri:2004sq} 
and also in the DSE-approach \cite{Schleifenbaum:2004id}, one obtains 
$\kappa = (93 - \sqrt{1201})/98 \approx 0.595$ 
\cite{Zwanziger:2001kw,Lerche:2002ep}, which satisfies the criterion. 
This specific value of $\kappa$ 
is found to vary only slightly for a large class of possible dressings 
of the ghost-gluon-vertex \cite{Lerche:2002ep}. Similar values have been 
determined 
from exact renormalisation group equations \cite{Pawlowski:2003hq}.

An interesting consequence of the solution (\ref{IRsolution}) is the 
qualitative universality of the running coupling in the infrared.  
Renormalisation group invariant
couplings can be defined from either of the primitively divergent vertices 
of Yang-Mills-theory, {\it i.e.} from the ghost-gluon vertex ($gh-gl$), 
the three-gluon vertex ($3g$) or the four-gluon vertex ($4g$) via
\beqa
\alpha^{gh-gl}(p^2) &=& \frac{g^2}{4 \pi} \, G^2(p^2) \, Z(p^2) 
     \hspace*{9mm} \stackrel{p^2 \rightarrow 0}{\sim} \hspace*{2mm} 
     {\bf const}/N_c \,, \label{gh-gl}\\
\alpha^{3g}(p^2) &=& \frac{g^2}{4 \pi} \, [\Gamma^{0,3}(p^2)]^2 \, Z^3(p^2) 
    \hspace*{2mm} \stackrel{p^2 \rightarrow 0}{\sim}
     \hspace*{2mm} {\bf const}/N_c \,,\\
\alpha^{4g}(p^2) &=& \frac{g^2}{4 \pi} \, [\Gamma^{0,4}(p^2)]^2 \, Z^4(p^2) 
    \hspace*{2mm} 
    \stackrel{p^2 \rightarrow 0}{\sim} \hspace*{2mm} {\bf const}/N_c \,.
     \label{alpha}
\eeqa
Using the DSE-solution (\ref{IRsolution}) it is easy to see that all three 
couplings approach a fixed point in the infrared. This fixed point can be
explicitly calculated for the coupling (\ref{alpha}). Employing a bare
ghost-gluon vertex one obtains $\alpha^{gh-gl}(0) \approx 8.92/N_c$
\cite{Lerche:2002ep}.

\section{Running coupling between Landau and Coulomb gauge}

One might expect that gauge invariant features of Yang-Mills
theory show up in a quantity like the running coupling. Thus it is  
interesting to determine the coupling together with the
infrared exponents of the corresponding dressing functions
also in other gauges. A particularly interesting class of 
(transverse) gauges is specified by the gauge condition
$a \partial_0 A_0 + {\bf \nabla \cdot A} = 0$ with the gauge parameter 
$0 \leq a \leq 1$.  The values $a=0,1$ correspond to Coulomb- and 
Landau-gauge respectively, whereas non-integer values of $a$ 
interpolate between these gauges.

In collaboration with D.~Zwanziger we showed that in these gauges
{\it two} renormalisation group invariant running couplings can be 
defined from the ghost-gluon vertex \cite{Fischer:2005qe}. 
In the Landau gauge limit, $a=1$, 
these couplings unify to the expression (\ref{gh-gl}). In the 
Coulomb gauge limit, however, they constitute two 
invariant charges which are given by
\beqa
\alpha_{\rm coul}(|{\bf k}|) &\equiv& 
     \ \frac{g^2}{4 \pi} \ Z_{00}^{\rm inst}( | { \bf k} | )\,, \\
\alpha_I(|{\bf k}|)  & \equiv &  \frac{16}{3} \ \frac{g^2}{4 \pi} 
      \ G^2(|{\bf k}|) \:  \frac{Z^{\rm tr}_{\rm et}(|{\bf k}|)}{|{\bf k}|} \,, 
\eeqa
where $Z_{00}^{\rm inst}$ is the instantaneous part of the time-time
component of the gluon dressing function and $Z^{\rm tr}_{\rm et}$
is the equal-time spatial gluon dressing function. The first coupling,
$\alpha_{\rm coul}$, is nothing else than the familiar colour
Coulomb potential, which has been shown to be (almost) linearly 
rising \cite{Zwanziger:2003de,Greensite:2003xf,Feuchter:2004mk}. 
The second coupling, $\alpha_I$, has a fixed point in the infrared \cite{Fischer:2005qe}. 
This can be seen from the Dyson-Schwinger equation of the ghost propagator in
Coulomb gauge. 
In interpolating gauges with $0 < a \leq 1$, this fixed point can be calculated 
and we find the same value $\alpha_I(0) \approx 8.92/N_c$ as in Landau gauge.
Correspondingly, the ghost and gluon propagator obey an infrared power law
with exponent $\kappa \approx 0.595$ independent of the gauge 
parameter $a$. The observation of these gauge independent features may
have the potential to bring 
us closer to a unified picture of confinement in Landau and Coulomb gauge.

\section{Numerical solutions for the propagators in the continuum and
 on a torus}

\begin{figure}[t]
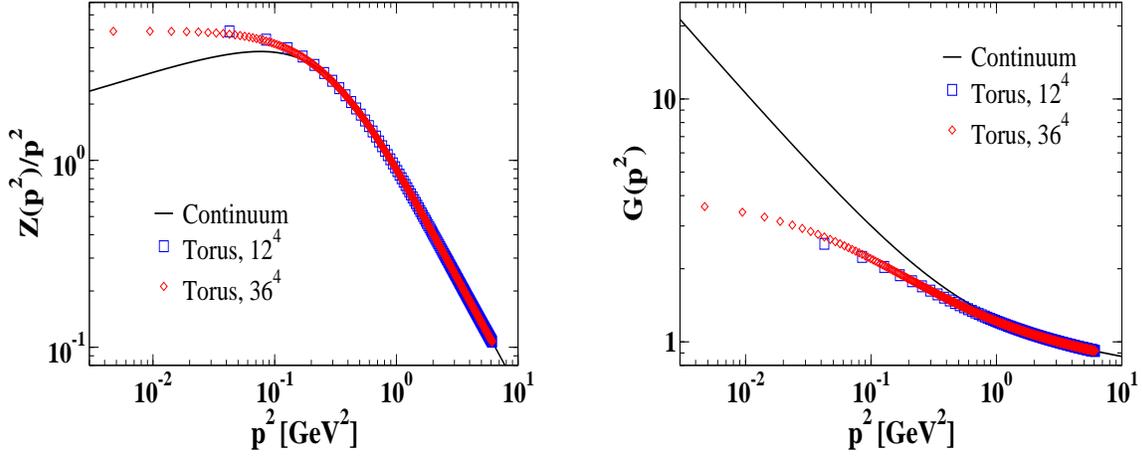

\centerline{
  \epsfig{file=volume.D.talk.eps,width=7cm,height=6cm}
  \hfill
  \epsfig{file=volume.G.talk.eps,width=7cm,height=6cm}}
\caption{The gluon propagator function $Z(p^2)/p^2$ (left diagram) and 
the ghost dressing function $G(p^2)$ (right diagram). Compared are the
results of the continuum DSEs with two respective solutions of DSEs on tori 
of different volumes.\label{vol1}}
\end{figure}
 
There are at least three caveats in comparing results from the continuum
Dyson-Schwinger approach to those of lattice calculations. First,
the quantitative aspects of the continuum solutions depend on the details
of the chosen truncation scheme, whereas the lattice calculations are 
{\it ab initio}. On the other hand, lattice calculations are carried out
on a compact manifold, and therefore one has to deal with effects due to 
finite volume and lattice spacing. Finally, boundary conditions have to
be chosen on the compact manifold. To quantify some of the latter two effects we
formulated the Dyson-Schwinger equations on a torus without changing the 
truncation scheme \cite{Torus,Fischer:2002hn}. One would then expect to 
see differences to the continuum solution for small volumes, which 
disappear continuously when the volume is chosen larger and larger.
However, for our solutions this is not the case. In fig.~\ref{vol1} we compare
solutions for the gluon propagator $Z(p^2)/p^2$ (left diagram) and the 
ghost dressing function $G(p^2)$ (right diagram) of the continuum
with those from two different tori with volumes of 
$V \approx 2500 \, \mathrm{fm}^4$ and $V \approx 10^5 \, \mathrm{fm}^4$. 
Note that the smaller volume corresponds to a typical one achieved in 
recent lattice simulations \cite{Bonnet:2001uh}, whereas the larger value 
exceeds the current possibilities of the lattice formulation by far.
Both solutions on the torus agree very well with each other, whereas
there is a clear difference to the continuum results. In terms of the
infrared exponent $\kappa$, defined in eq. (\ref{kappa}), one finds a value
of $\kappa \approx 0.5$ for the gluon propagator 
on the torus, which should be compared to the corresponding
continuum value $\kappa \approx 0.6$. Similarly, whereas the ghost dressing
function diverges in the continuum, we observe a finite ghost on the
compact manifold. These differences persist, when the infrared behaviour 
on the torus is extrapolated to the continuum \cite{Torus}. 

\begin{figure}[t]
\centerline{
  \epsfig{file=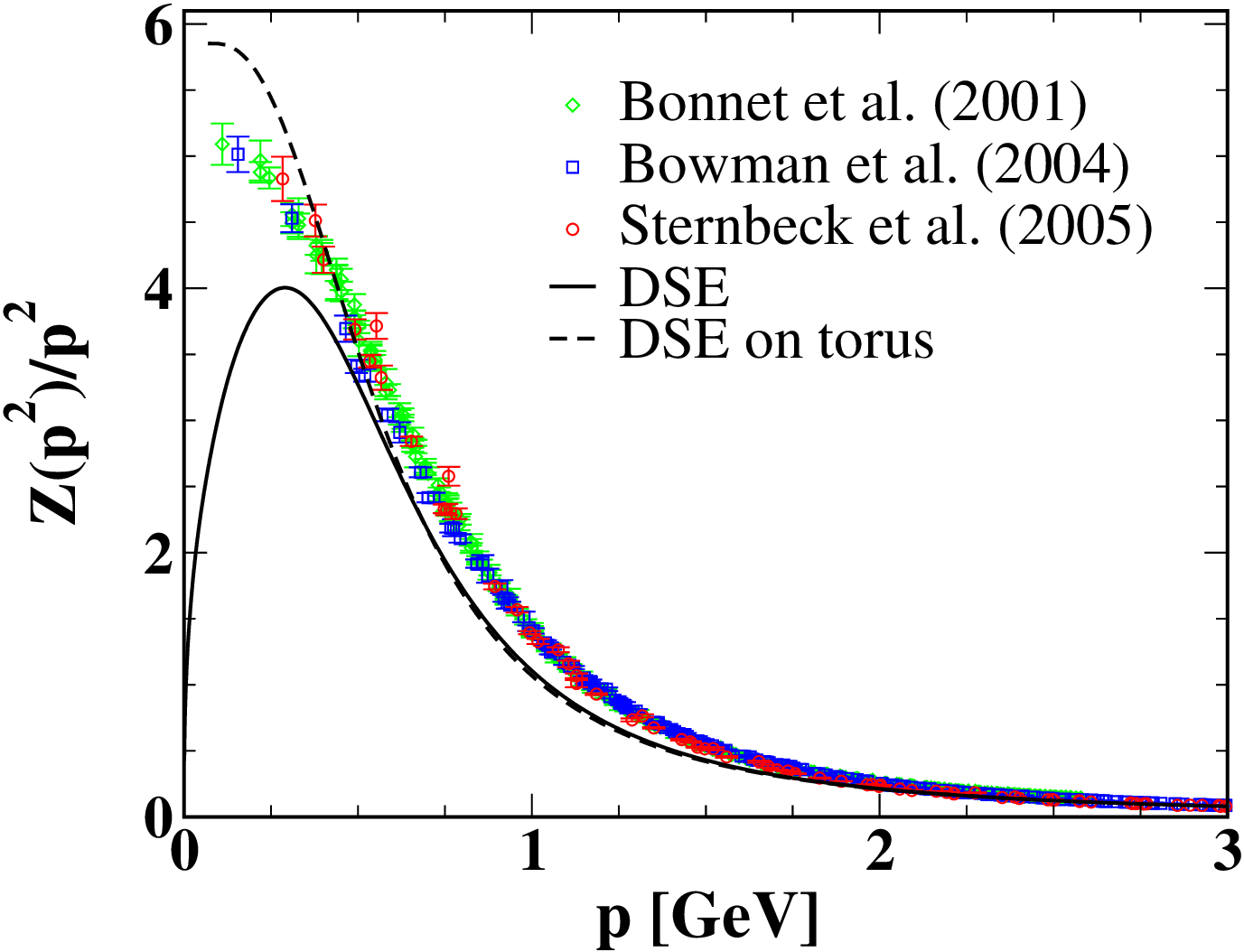,width=7cm,height=6cm}
  \hfill
  \epsfig{file=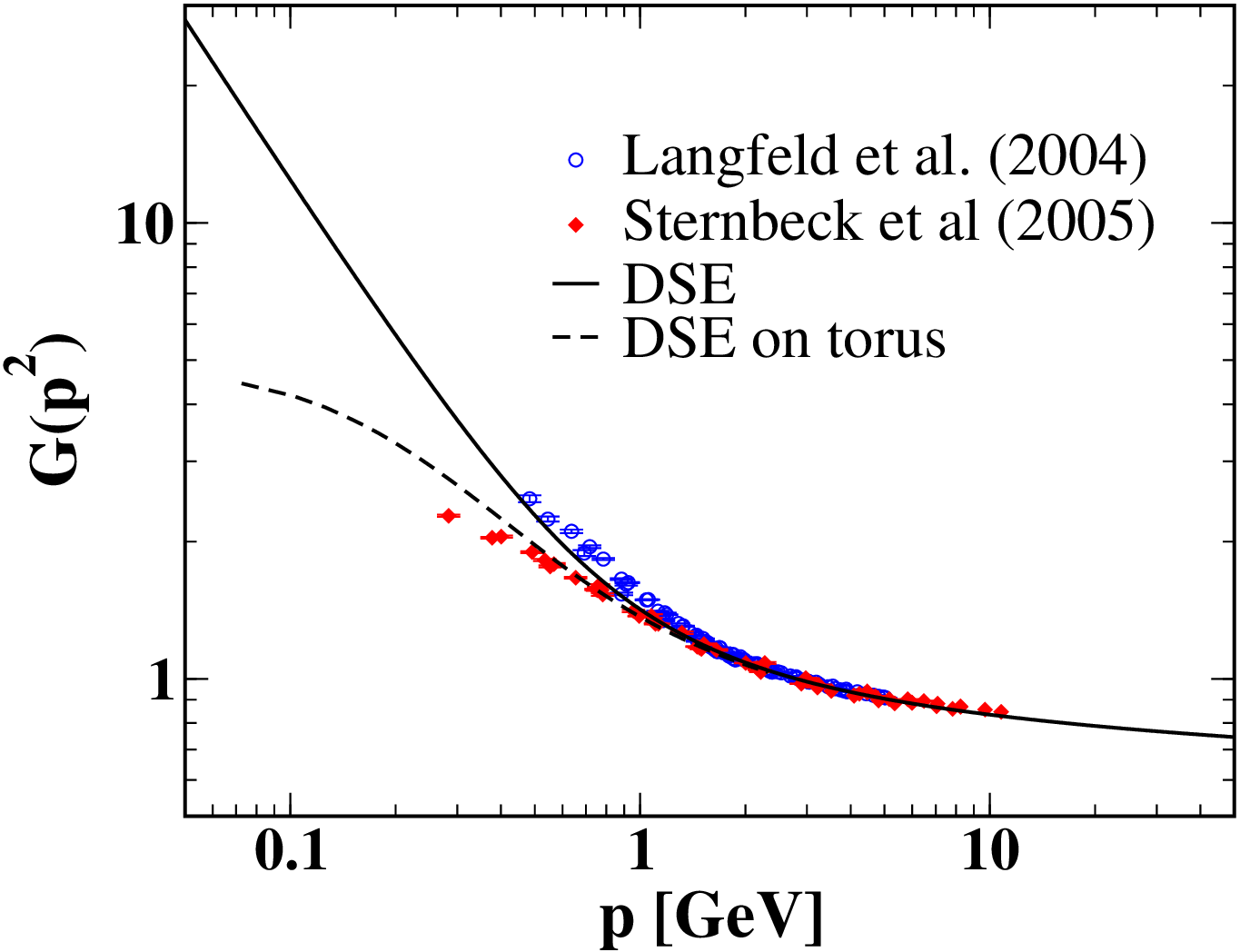,width=7cm,height=6cm}}
\caption{The results for gluon and ghost from Dyson-Schwinger equations 
in the continuum and on the torus are compared with the lattice data of
refs.~\cite{Bonnet:2001uh,Bowman:2004jm,Sternbeck:2005tk,Gattnar:2004bf}\label{vol2}}
\vspace*{8mm}
\centerline{
  \epsfig{file=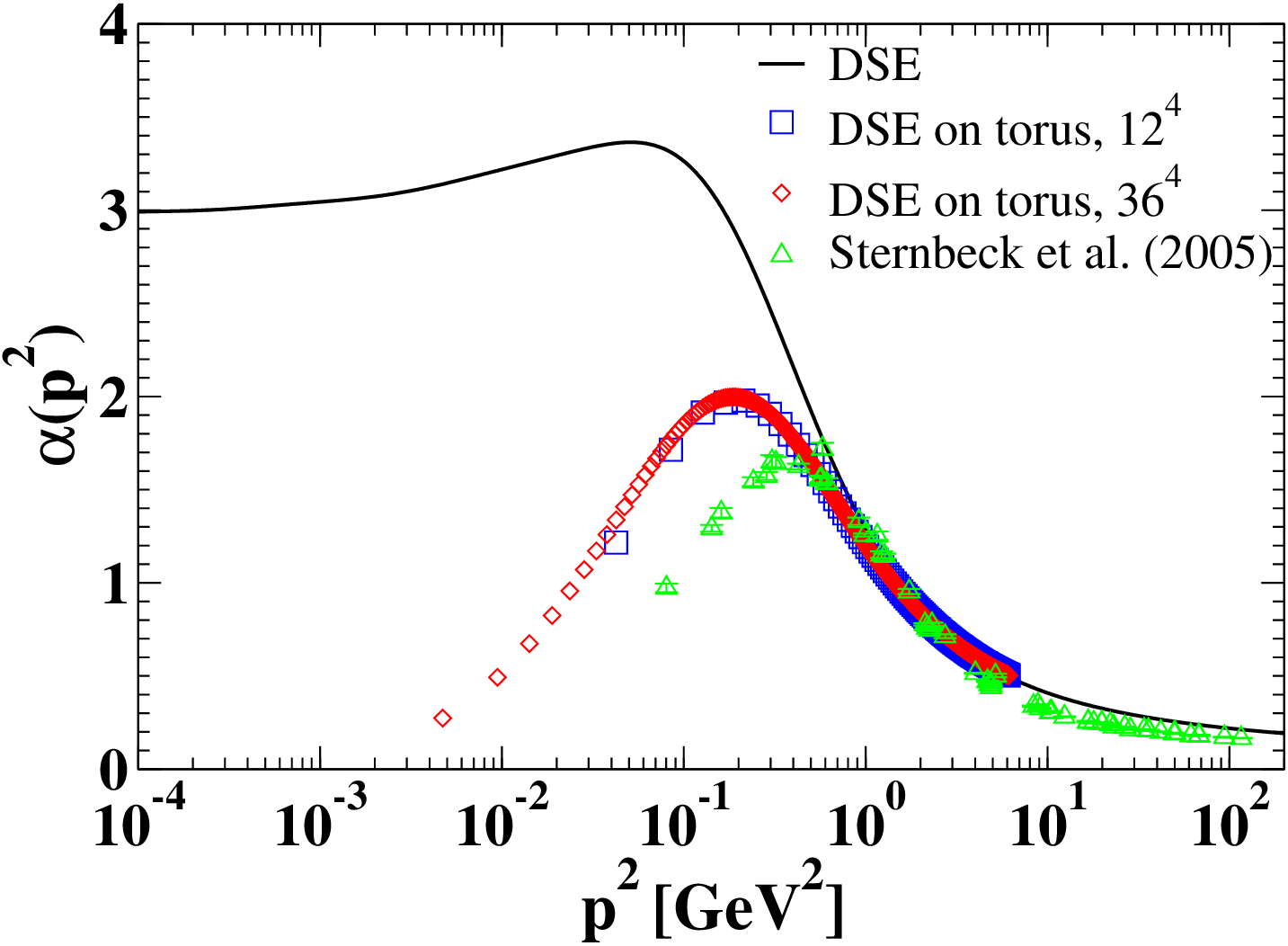,width=8cm}}
\caption{Results for the running coupling of the ghost-gluon vertex from
DSEs in the continuum and on the torus compared to the lattice data
of ref.~\cite{Sternbeck:2005tk}.\label{vol3}} 
\end{figure}

We compare these results to the ones of recent lattice simulations in
fig.~\ref{vol2} and fig.~\ref{vol3}. The lattice simulations for the
gluon propagator are all performed for SU(3) at different values of $\beta$, 
different lattice sizes and different actions 
\cite{Bonnet:2001uh,Bowman:2004jm,Sternbeck:2005tk} (see also \cite{Oliveira:2004gy}).
They fall on top of each other. On the other hand there are slight 
differences between the SU(2) ghost dressing function
of ref.~\cite{Gattnar:2004bf} and the SU(3) result of ref.~\cite{Sternbeck:2005tk}
(see also \cite{Furui:2003jr}).
Interestingly, there is nice qualitative agreement between both, 
the lattice and the DSE-formulation on the compact manifold. Both 
approaches give a finite gluon propagator in the infrared. 
This is in marked difference to the numerical continuum result, 
which agrees with the analytic power law, eq.(\ref{kappa}), with
$\kappa \approx 0.6$. The situation
is less clear for the ghost dressing function, though at least the
SU(3) result of ref.~\cite{Sternbeck:2005tk} seems to follow more or less
the DSE-solution on the torus. For the running coupling 
$\alpha^{gh-gl}(p^2) = \frac{g^2}{4 \pi} G^2(p^2) Z(p^2)$ the numerical
solution in the continuum reproduces the fixed point 
$\alpha^{gh-gl}(0) \approx 8.92/N_c$ 
discussed below eq.(\ref{alpha}). On the compact manifold, however, 
the coupling 
vanishes at small momenta\footnote{A similar result has been found 
for the running coupling of the 
three-gluon vertex in refs.~\cite{Alles:1996ka,Boucaud:2005qf}. 
Note, however, that the conclusions drawn in \cite{Boucaud:2005qf}
are contradicted by the results of 
refs.~\cite{Alkofer:2004it,Watson:2001yv,Cucchieri:2004sq,Lerche:2002ep}.
}, a behaviour which seems to be at least counterintuitive.

Our results from DSEs compared to lattice
simulations of the ghost and gluon propagators show
an interesting difference between the infrared behaviour found in the 
continuum and the one found on a compact manifold.
Finite volume effects can hardly explain this result, since
the difference persists for extremely large tori. The role of different
boundary conditions in this respect is currently investigated 
({\it c.f.} also ref.~\cite{tok}, these proceedings). 

\pagebreak

{\bf Acknowledgements}

I would like to thank the organisers of {\it Lattice 2005\/}
for all their efforts which made this interesting conference possible.
I am grateful to R.~Alkofer, B.~Gruter, F.~Llanes-Estrada and 
D.~Zwanziger for fruitful collaboration on the topics presented here. 
I thank D.~Leinweber, O.~Oliveira, M.~Pennington, J.~Skullerud and 
A.~Williams for inspiring discussions.  
This work has been supported by the Deutsche For\-schungsgemeinschaft
(DFG) under contract Fi 970/2-1.


\begin{thebibliography}{99}

\bibitem{Nakanishi:qmas}
N.~Nakanishi and I.~Ojima,
World Sci.\ Lect.\ Notes Phys.\  {\bf 27}, 1 (1990).

\bibitem{Alkofer:2000wg}
R.~Alkofer and L.~von Smekal,
Phys.\ Rept.\  {\bf 353}, 281 (2001).

\bibitem{Zwanziger:2003de}
D.~Zwanziger,
Phys.\ Rev.\ D {\bf 70} (2004) 094034.

\bibitem{Alkofer:2004it}
R.~Alkofer, C.~S.~Fischer and F.~J.~Llanes-Estrada,
Phys.\ Lett.\ B {\bf 611}, 279 (2005).
  
\bibitem{Fischer:2005qe}
C.~S.~Fischer and D.~Zwanziger,
Phys.\ Rev.\ D {\bf 72} (2005) 054005.

\bibitem{Torus} 
C.~S.~Fischer, B.~Gruter and R.~Alkofer, arXiv:hep-ph/0506053.

\bibitem{Watson:2001yv}
P.~Watson and R.~Alkofer,
Phys.\ Rev.\ Lett.\  {\bf 86} (2001) 5239.

\bibitem{Cucchieri:2004sq}
A.~Cucchieri, T.~Mendes and A.~Mihara,
JHEP {\bf 0412} (2004) 012.

\bibitem{Schleifenbaum:2004id}
W.~Schleifenbaum, A.~Maas, J.~Wambach and R.~Alkofer,
Phys.\ Rev.\ D {\bf 72} (2005) 014017.

\bibitem{Zwanziger:2001kw}
D.~Zwanziger,
Phys.\ Rev.\ D {\bf 65},  094039 (2002).

\bibitem{Lerche:2002ep}
C.~Lerche and L.~von Smekal,
Phys.\ Rev.\ D {\bf 65}, 125006 (2002).

\bibitem{Pawlowski:2003hq}
J.~M.~Pawlowski, D.~F.~Litim, S.~Nedelko and L.~von Smekal,
Phys.\ Rev.\ Lett.\  {\bf 93} (2004) 152002
;\\
C.~S.~Fischer and H.~Gies,
JHEP {\bf 0410} (2004) 048,

\bibitem{Greensite:2003xf}
J.~Greensite and S.~Olejnik,
Phys.\ Rev.\ D {\bf 67} (2003) 094503.

\bibitem{Feuchter:2004mk}
C.~Feuchter and H.~Reinhardt,
Phys.\ Rev.\ D {\bf 70} (2004) 105021;
C.~Feuchter and H.~Reinhardt,
arXiv:hep-th/0402106.

\bibitem{Fischer:2002hn}
C.~S.~Fischer  and R.~Alkofer,
Phys. Lett. B {\bf 536}, 177 (2002);\\
C.~S.~Fischer, R.~Alkofer and H.~Reinhardt,
Phys.\ Rev.\ D {\bf 65}, 094008 (2002).

\bibitem{Bonnet:2001uh}
F.~D.~Bonnet {\it et al.},
Phys.\ Rev.\ D {\bf 64}, 034501 (2001).

\bibitem{Bowman:2004jm}
P.~O.~Bowman, U.~M.~Heller, D.~B.~Leinweber, M.~B.~Parappilly and 
A.~G.~Williams,
Phys.\ Rev.\ D {\bf 70} (2004) 034509.

\bibitem{Sternbeck:2005tk}
A.~Sternbeck, E.~M.~Ilgenfritz, M.~Mueller-Preussker and A.~Schiller,
Phys.\ Rev.\ D {\bf 72} (2005) 014507.

\bibitem{Oliveira:2004gy}
O.~Oliveira and P.~J.~Silva,
AIP Conf.\ Proc.\  {\bf 756}, 290 (2005).

\bibitem{Gattnar:2004bf}
J.~Gattnar, K.~Langfeld and H.~Reinhardt,
Phys.\ Rev.\ Lett.\  {\bf 93}, 061601 (2004).

\bibitem{Furui:2003jr}
S.~Furui and H.~Nakajima,
Phys.\ Rev.\ D {\bf 69} (2004) 074505.

\bibitem{Alles:1996ka}
B.~Alles, D.~Henty, H.~Panagopoulos, C.~Parrinello, C.~Pittori and D.~G.~Richards,
Nucl.\ Phys.\ B {\bf 502} (1997) 325.

\bibitem{Boucaud:2005qf}
P.~Boucaud {\it et al.},
JHEP {\bf 0304}, 005 (2003);
P.~Boucaud, F.~De Soto, A.~Le Yaouanc, J.~P.~Leroy, J.~Micheli, 
O.~Pene and J.~Rodriguez-Quintero,
arXiv:hep-ph/0505150.

\bibitem{tok}
T.~Tok, L.~v.~Smekal, K.~Langfeld, H.~Reinhardt, 
``The gluon propagator in lattice Landau gauge with 
twisted boundary conditions,'',these proceedings. 


  
\end{thebibliography}
\end{document}